\documentclass[]{raa} 
\usepackage{graphicx,times}             
\usepackage{amsmath}
\usepackage{threeparttable}
\usepackage{graphicx}
\usepackage{longtable}
\usepackage{natbib}
\bibpunct{(}{)}{;}{a}{}{,}

\begin{document}

   \title{An Isolated Compact Galaxy Triplet}
   \volnopage{Vol.0 (200x) No.0, 000--000}      
   \setcounter{page}{1}          

   \author{Shuai Feng
      \inst{1,2}
   \and Zheng-Yi Shao
      \inst{1 \dag}
   \and Shi-Yin Shen
      \inst{1,4}
   \and Maria Argudo-Fern$\mathrm{\acute a}$ndez
      \inst{1}
   \and Hong Wu
      \inst{3}
   \and Man-I Lam
      \inst{3}
   \and Ming Yang
      \inst{3}
   \and Fang-Ting Yuan
      \inst{1}
   }
   \institute{Key Laboratory for Research in Galaxies and Cosmology, Shanghai Astronomical Observatory, Chinese Academy of Sciences, 80 Nandan Road, Shanghai 200030, China. {Email: zyshao@shao.ac.cn}{\dag}\\
        \and
             University of the Chinese Academy of Sciences, No.19A Yuquan Road, Beijing 100049, China \\
        \and
             Key Laboratory of Optical Astronomy, National Astronomical Observatories, Chinese Academy of Sciences, 20A Datun Road, Beijing 100012, China \\
        \and Key Lab for Astrophysics, Shanghai, 200234, China\\
  }

   \date{Received~~2009 month day; accepted~~2009~~month day}

\abstract{We report the discovery of an isolated compact galaxy triplet
SDSS J084843.45+164417.3, which is first detected by the LAMOST spectral survey and
then confirmed by the spectroscopic observation of the BFOSC of the 2.16 meter telescope.
It is found that this triplet is  an isolated and extremely compact system, which has
an aligned configuration and very small radial velocity dispersion.
The member galaxies have similar colors and show marginal star formation
activities. These results enhance the opinion that the compact triplets are
well-evolved systems rather than the hierarchically forming
structures. This occasional discovery reveals the limitations of the
fiber spectral redshift surveys in studying such compact system, and declares
the necessity of  additional observations to complete the
current redshift sample. }

   \authorrunning{S. Feng, Z.-Y. Shao, S.-Y. Shen et al. }  
   \titlerunning{Discovery of an Isolated Galaxy Triplet }  

   \maketitle

\section{INTRODUCTION}

Galaxy triplet constitutes the simplest galaxy group and the
smallest N-body system that cannot be modelled analytically. In
general, triplets have not received much attention in contrast to
the close pairs or the rich groups of galaxies. So, the basic
scenario of their formation and evolution is still ambiguous.
According to the popular hierarchical structure formation theory,
it is naturally presumed that a triplet is formed through a
close galaxy pair accreting the third remote galaxy. However,
some recent evidences suggest that for lots of triplets, their
member galaxies have similar properties and are supposed to be
the remains of a long evolved dynamical system within a
common dark matter halo ( e.g. \citealt{Duplancic2015} and \citealt{Aceves2001}).

The first catalogue of galaxy triplets was compiled by
\citet{Karachentseva1979} and \citet{Karachentsev1988}, which
contains 84 Northern isolated galaxy triplets. These triplets were
selected by visual inspection of Palomar Sky Survey plates, whose
member galaxies have apparent magnitudes brighter than 15.7. About
64 percent of these targets were considered to be physical triplets
with $\Delta v_{ij}<500 kms^{-1}$ (\citealt{Karachentsev1981},
hereafter K-sample).  Based on this catalogue, the basic
properties of the galaxy triplets, e.g. integrated luminosity,
diameter, velocity dispersion, as well as the spatial
configuration, dynamics and dark matter content have been
estimated and discussed in detail (\citealt{Karachentseva1982};
\citealt{Karachentseva1983}; \citealt{karachentsev1990};
\citealt{Chernin1991}; \citealt{Anosova1992}; \citealt{Zheng1993}
and \citealt{Aceves2001}) .  For the Southern sky
($\delta<3^{\circ}$), similar works have also been done on a
sample of 76 isolated triplets, which are selected from the
European Southern Observatory/Science and Engineering Research
Council (ESO/SERC) and Palomar Observatory Sky Survey first
release (POSS-I) \citep{Karachentseva2000}.

\citet{omill2012} constructed a catalogue of triplets with their
primary galaxies ($Mr < -20.5$) selected from a volume-limited
sample in the Sloan Digital Sky Survey Data Release 7 (SDSS-DR7)
\cite{Abazajian2009}(hereafter O-sample). This catalogue comprises
1092 triplets with redshift range $0.01 \leq z \leq 0.14$.
However, due to the fiber collision effect of the SDSS (minimum
separation of 55'' between any two fibers), most of the companion
galaxies in these systems were selected by only using their
photometric redshifts ($z_{phot}$). As a result of the relatively
large uncertainty of $z_{phot}$ ($\sim 0.0227$), this catalogue is
inevitable contaminated by "projected triplets" and yet, in
another respect, some of the real triplets could be missed.  In
O-sample, there are about one tenth of the triplets with
spectroscopic redshifts for all their members.
\citet{Duplancic2013,Duplancic2015}  analyzed the configuration
and dynamics of this sub-sample, and compared them with galaxy
pairs and clusters. They found that the triplet galaxy members are
more similar to the galaxies in compact groups and rich clusters
than in galaxy pairs and concluded  that the galaxy triplets may
not be formed hierarchically. Very recently, \citet{Argudo2015}
published a sample of 315 isolated triplets in the local
Universe($z\leq 0.080$) using the spectroscopic redshifts
 ($z_{spec}$) from SDSS DR 10 (hereafter A-sample). In this sample,
galaxies are considered to be physically bound  to the primary
galaxy at projected separation up to $d\le$450 kpc and with
radial velocity difference $\Delta v \le 160 {\rm km s^{-1}}$.

Given the limited data and studies summarized above, there are
still not enough pieces to solve the puzzles  of the formation and
evolution scenario of the galaxy triplets. Any additional sample
of physical triplets, especially for those compact ones, will be
valuable to accumulate our knowledge about this kind of system.

\begin{figure}
   \centering
\includegraphics[width=0.4\textwidth]{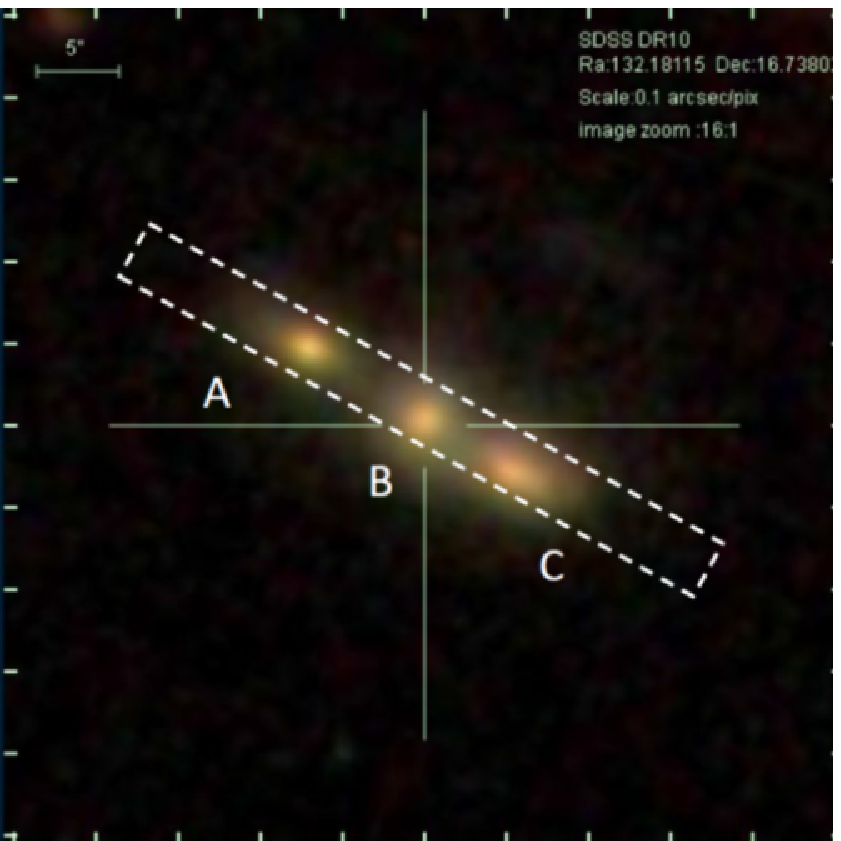}
\includegraphics[width=0.4\textwidth]{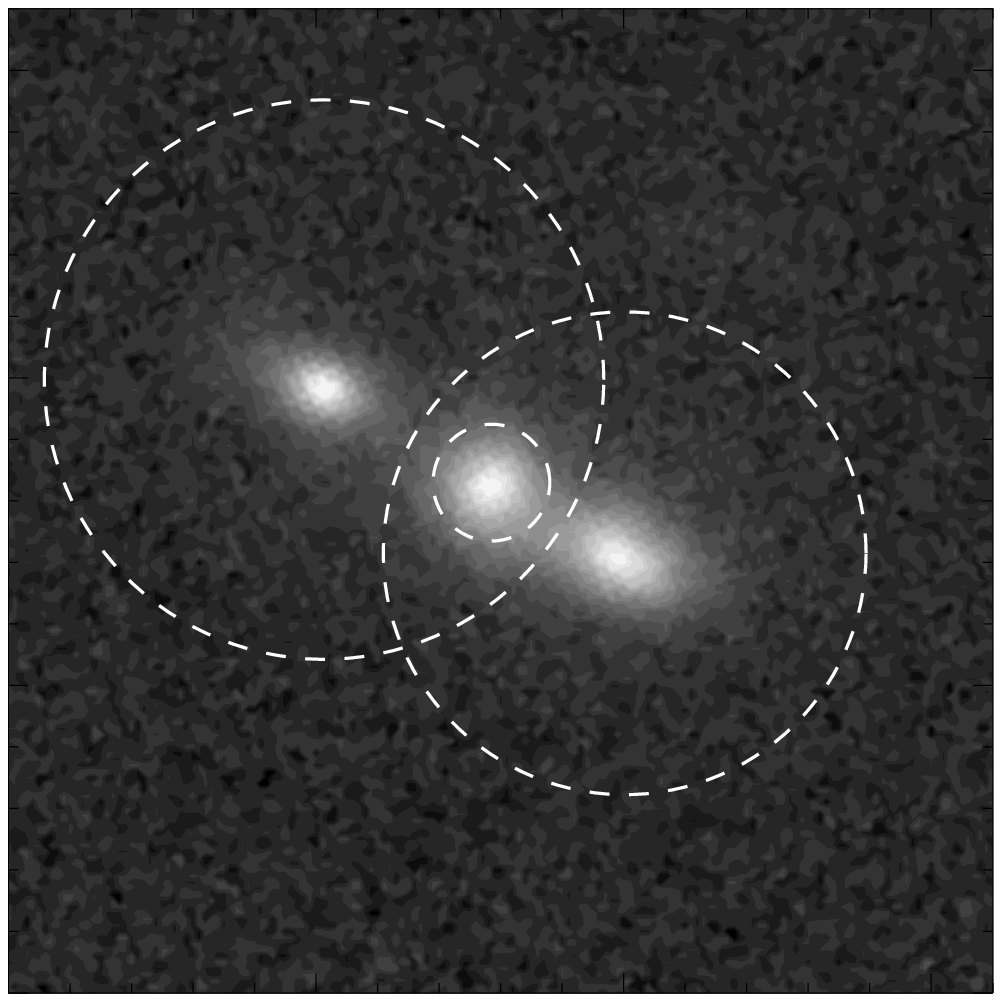}
\includegraphics[width=0.4\textwidth]{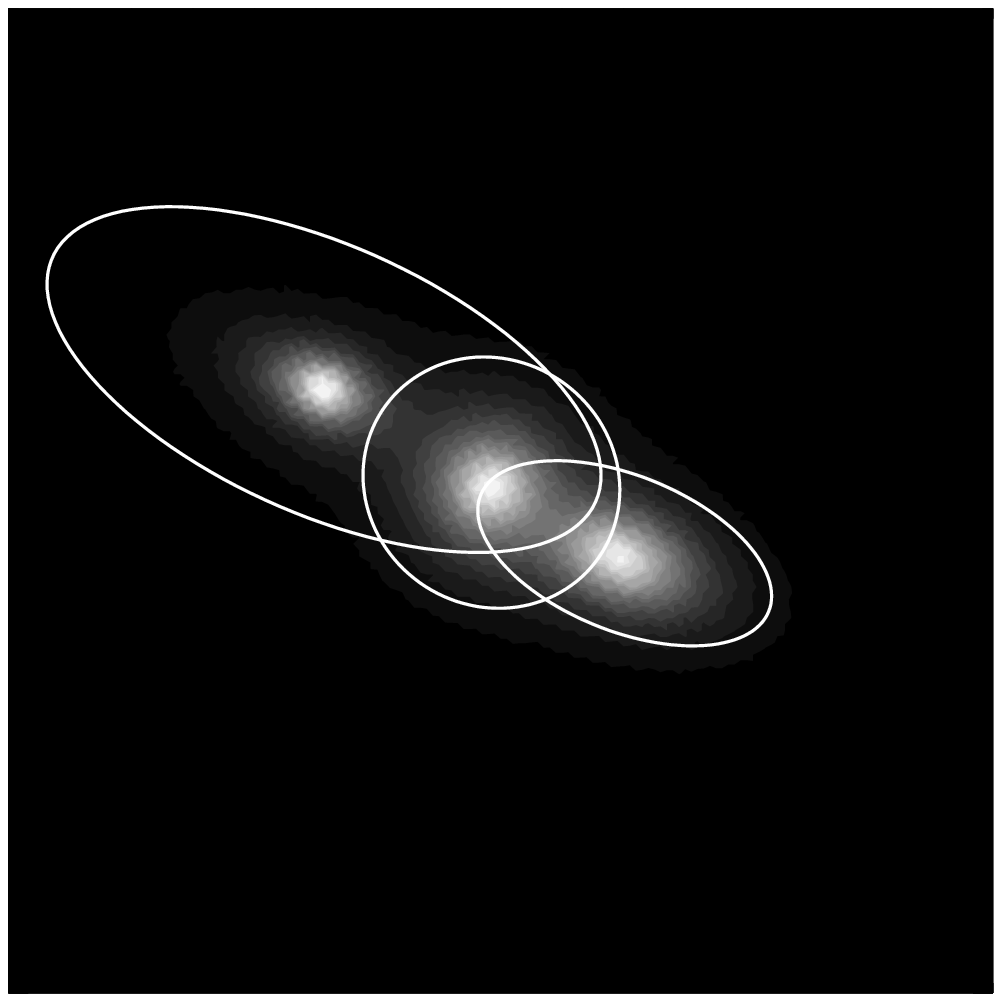}
\includegraphics[width=0.4\textwidth]{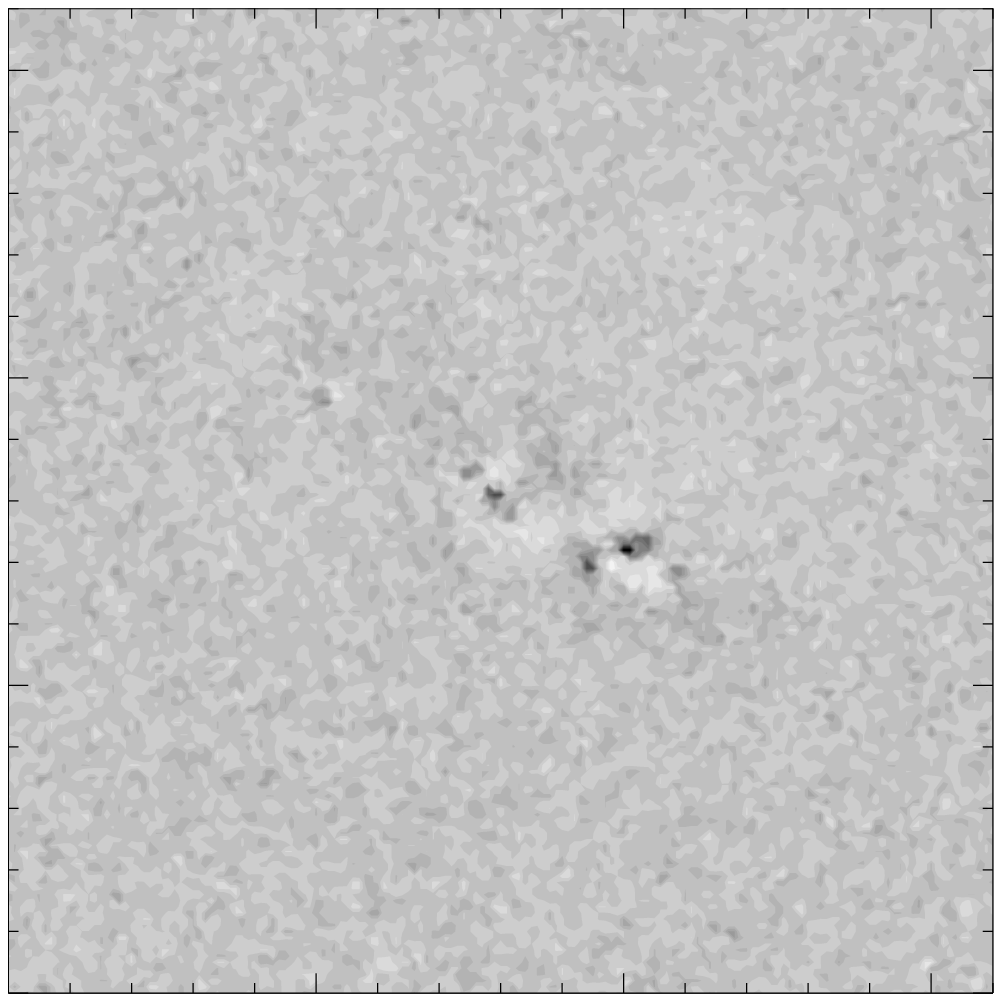}
\caption{ Images of J0848+1644. \emph{Top Left:} The optical image
combined with SDSS $g,r,i$ bands. The letters (A,B and C) labelled
in the figure are used to denote the members of the triplet. The
white dashed cube indicates the position of the long slit of the
BFOSC of the 2.16m telescope. The 5$^{ \prime\prime}$ scale line
at the up-left corner corresponds to 8.2kpc at the redshift of
this triplet. \emph{Top Right:} The $r$-band SDSS frame image
where the dashed circles indicate the Petrosian $R_{90}$ from
SDSS. \emph{Bottom Left:} The best fitting model made by
\texttt{GALFIT} where the solid circles indicate the isophote
ellipses enclose 90\% total model flux of individual galaxies.
\emph{Bottom Right:} Residual image of the best fitting, the grey
scale from black to white indicates -5\% to 8\% of the maximum
flux in the model image. }
   \label{Fig:plot1}
\end{figure}

The galaxy system SDSS J084843.45+164417.3 (hereafter J0848+1644),
which contains galaxy A, B and C (see Fig. \ref{Fig:plot1} for
details), is the spectroscopic triplet we report and study in this
paper. This triplet is not included in the A-sample because these
three galaxies have very small angular distances and only galaxy B
has spectrum in SDSS, $z_{spec}=0.078829\pm0.000015$ (
SpecObjID: 2565950139160094720) \citep{Abazajian2009}.  Also, it
is excluded by the O-sample since the $z_{phot}$ of the other two
galaxies ($z_{phot}=(0.32, 0.08,
0.19)$ for galaxy A, B and C respectively) are all much more
different from $z_{spec}$ of galaxy B. This galaxy triplet was reported as
a galaxy pair in \citet{Shen2015}, where they started a project
that aims to get the spectroscopic redshifts of all the main
sample of galaxies($r<17.77$) yet without redshifts measured in
SDSS DR7 due to the fiber collision. Their new redshift survey is
operated on Guo Shou Jing telescope \citep[also known as Large
Area Multiple Objects Spectroscopic Telescope Survey, hereafter
LAMOST,][]{Cui2012}. The galaxy C is compiled into the
complementary galaxy sample of the LAMOST spectral survey and its
spectra has been released in the LAMOST DR1 with $z_{spec}$ = 0.07929 (Obsid:78907168)
\citep{Luo2015}\footnote{Based on SDSS DR7 photometry, the galaxy
C ($r$=17.06) belongs to the main galaxy sample whereas the galaxy A
($r$=18.45) does not.}. During the visual inspection of the galaxy
pairs in \cite{Shen2015}, we find, together with galaxy A, they
are very possibly in a compact triplet system.

We have applied a follow-up spectroscopic confirmation of this
triplet using the 2.16 meter optical telescope (hereafter 2.16m)
on the Xinglong Station of the National Astronomical Observatories
of Chinese Academy of Sciences. In this paper, we firstly report
the data from the spectroscopic observation on 2.16m. After that,
we combine  all optical photometric and spectroscopic data and
make a detailed study on this triplet. This paper is organized as
follows. In section 2, we describe the photometric and
spectroscopic data of J0848+1644 and the measurements of the basic
features of its members. In section 3, we derive the global
properties of this triplet, such as the compactness,
configuration, environment, dynamical status and star forming rate
etc. Finally, we present discussions in section 4 and list summary
in section 5. Throughout this paper, we use cosmological
parameters $\Omega_\Lambda$ = 0.7, $\Omega_M$= 0.3, and h = 0.7.

\section{OBSERVATIONS AND DATA REDUCTION} \label{obsdat}

In this section, we describe the observation of triplet
J0848+1644 from 2.16m telescope and our new photometric data reduction on the SDSS images.
All measurements of individual member galaxies are summarized
in Table \ref{tabel1}.

\begin{figure}
\centering
\includegraphics[width=1.05\textwidth]{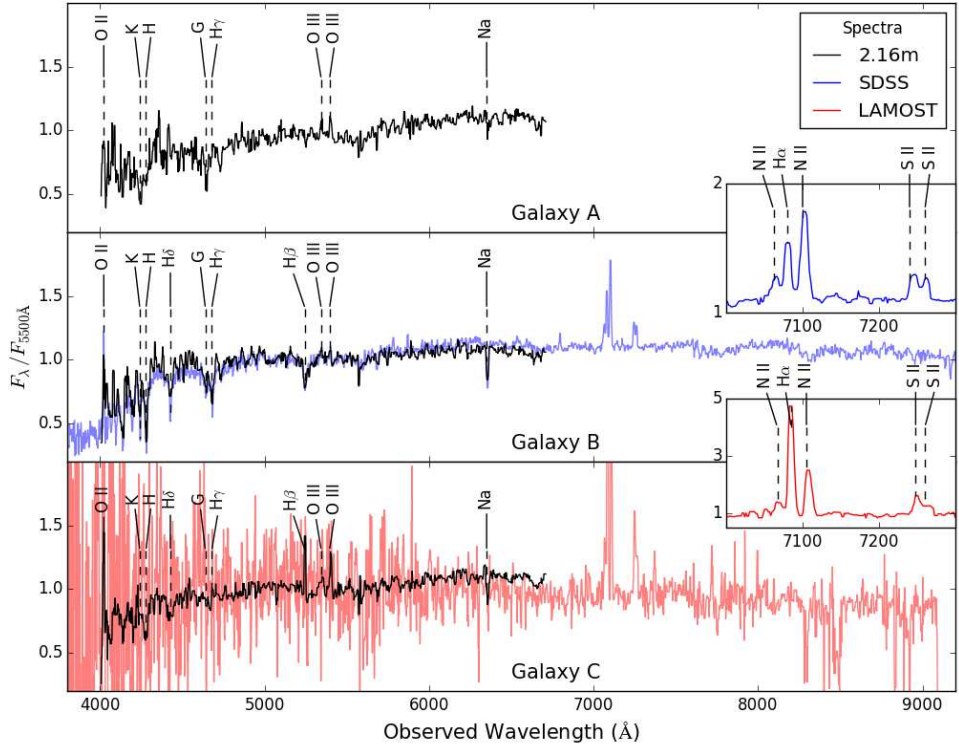}
\caption{The optical spectra of member galaxies of the J0848+1644.
The black lines are the spectra of the BFOSC of the 2.16m
telescope. The blue and red lines are taken from SDSS for galaxy B
and LAMOST for galaxy C respectively. All the spectra are
normalized at 5500 \text{\AA} and smoothed with a width of 11
\text{\AA}. The redshifts of individual spectra are estimated by
averaging the measurements of multiple spectral lines, whose
species names are labelled with corresponding spectra in three
main panels. Additionally, we show some spectral lines of SDSS and
LAMOST spectra in minor panel which are not covered by 2.16m telescope.}
   \label{Fig:plot2}
\end{figure}

\begin{table}
\begin{center}
  \caption[]{Observational and derived parameters of individual members of J0848+1644}\label{tabel1}
  \smallskip
  \setlength{\tabcolsep}{9pt}
  \begin{threeparttable}
  \newcommand{\tabincell}[2]{\begin{tabular}{@{}#1@{}}#2\end{tabular}}
  \begin{tabular}{llrrr}
  \hline\noalign{\smallskip}
  Parameter &  Unit  &   galaxy A     &   galaxy B     &   galaxy C     \\
  \hline\noalign{\smallskip}
  RA (J2000) & hh:mm:ss.ss & \tabincell{r@{:}l}{08&48:43.94}    & \tabincell{r@{:}l}{08&48:43.45}    & \tabincell{r@{:}l}{08&48:43.08}      \\
  DEC(J2000) & dd:mm:ss.ss & \tabincell{r@{:}l}{$+$16&44:21.76} & \tabincell{r@{:}l}{$+$16&44:17.36} & \tabincell{r@{:}l}{$+$16&44:14.18}   \\
  $d_{i}$   \tnote{a}   &   kpc       &   13.97            &   -                &   10.44            \\
  \hline\noalign{\smallskip}
  $z_{SDSS}$  &       &       & 0.078829$\pm$0.000015 &       \\
  $z_{LAMOST}$ \tnote{b} &       &       &       & 0.07929  \\
  \hline\noalign{\smallskip}
  $z$ &     & 0.07852$\pm$0.00015 & 0.07848$\pm$0.00016 & 0.07883$\pm$0.00017 \\
  $v_r$      &  km s$^{-1}$       & 23557$\pm$45 & 23544$\pm$48 & 23650$\pm$51 \\
  $F$(OII)  &    10$^{-17}$ erg s$^{-1}$ cm$^{-2}$    &  200.4  &  453.6  &  1396.1   \\
  \hline\noalign{\smallskip}
  $m_r$ \tnote{c}   &           &   17.32   &   17.00   &   16.99   \\
  $n_{s}$ \tnote{c}      &      &   6.0            &   2.1            &   1.6            \\
  $R_e$ \tnote{c}  &  kpc      &   3.6            &   2.6            &   3.6            \\
  $b/a$   \tnote{c}   &           &   0.45    &   0.94    &   0.50    \\
  $\theta$  \tnote{c}   &  degree   &   5.3   &   -5.2  &   7.6   \\
  $R_{a,90}$  \tnote{d} &  kpc      &   21.0            &   9.0            &   11.0            \\
  $R_{c,90}$  \tnote{e} &  kpc      &   14.9            &   8.9            &   8.3            \\
  \hline\noalign{\smallskip}
  \multicolumn{2}{l}{$g-r$  } &  0.75  &  0.77  & 0.78  \\
  $\log(M_{\ast}/M_{\odot})$ \tnote{f} &        &   10.51   &   10.66   &   10.67   \\
  $SFR$ \tnote{g} &  M$_{\odot}\cdot$ yr$^{-1}$  &  0.38      &  0.86     &  2.64       \\
  $\log(SFR/M_{\ast})$    &       & -10.9      & -10.7     & -10.3       \\
  \noalign{\smallskip}\hline
  \end{tabular}
       \begin{tablenotes}
        \item[a] Projected distance from the central galaxy B.
        \item[b] The redshift error of galaxy C is not provided in LAMOST DR1.
        \item[c] Fitting parameters that derived from the \texttt{GALFIT} by modelling the $r$-band image with three Sersic profile components.
        \item[d] Major axis radius of ellipse enclosing 90\% model flux in the $r$-band.
        \item[e] Radii of circle enclosing 90\% model flux in the $r$-band.
        \item[f] Stellar mass estimated according to \citet{bell2013}.
        \item[g] Estimated from F(OII).
       \end{tablenotes}
  \end{threeparttable}
\end{center}
\end{table}

\subsection{Spectroscopy}

\subsubsection{Observations}

Our follow-up spectroscopic observation was taken by BAO Faint
Object Spectrograph and Camera (hereafter BFOSC) of the 2.16m
optical telescope in Dec. 31, 2013. The long slit of BFOSC is
1.8$^{\prime\prime}$ wide and covers all three galaxies
simultaneously (as shown in Fig. \ref{Fig:plot1}). The spectral
wavelength coverage is from 3870 to 6760 \text{\AA}, and the
resolution is $R \sim$ 700 at 5000 \text{\AA}.

The spectroscopic data were reduced following the standard
procedures using the NOAO Image Reduction and Analysis Facility
(IRAF, version 2.16) software package, including the bias and
flat-field correction of CCD, and the cosmic-ray removal.
Wavelength calibration was performed by comparing with the
ferrum/argon lamp spectra, which exposed at both the beginning and
the end of that observation. Flux calibration of all spectra was
performed using the additional observation of the standard
star, Feige 34. The atmospheric extinction was corrected
with the mean extinction coefficients of the Xinglong station.

The resultant 1-dimensional spectra have typical $S/N \sim 14$.
For galaxy B, we plot  the spectrum of SDSS (see Fig.
\ref{Fig:plot2}) for comparison. These two spectra matched very
well in their overlap wavelength range, which also demonstrates the
reasonability of our flux calibration of BFOSC.

\subsubsection{Radial velocities and emission line strengths}

To get the redshift measurements from the  BFOSC spectra, we
firstly measure the radial velocities of each individual emission
or absorbtion line ($z_l$) by using the IRAF package
\texttt{splot}. The redshift of each galaxy is then calculated by
averaging the results of all its adopted spectral lines,
$z=\langle z_{l} \rangle$, and the error is estimated from their
dispersion, $z_{err}= \sigma_{z_{l}} / \sqrt {N_{l}}$. Some of the
identified lines are not strong enough or blended by sky lines,
and therefore have not been used in the redshift calculation. The
final adopted spectral lines for each spectrum are labeled in
main panels of Fig. \ref{Fig:plot2} respectively and the redshift results are
listed in Table \ref{tabel1}. For galaxy B, the BOSFC redshift is
in excellent agreement with the SDSS value ($z_{SDSS}=0.078829$).
For galaxy C, the BOSFC redshift is also consistent with the
LAMOST value ($z_{LAMOST}=0.0793$). Considering the internal
consistence of the BOSFC spectra, we only take the BOSFC redshifts
to study the kinematics and dynamics of J0848+1644 below.

Besides the redshifts, the  fluxes of OII emission lines of three
BFOCS spectra have also been measured through the Gaussian fit and
the correction of the long slit effect, which will be used as
diagnostics of the current star formation (SFR) of each galaxy.


\subsection{Photometric measurements}

The SDSS catalogue has released a variety of photometric
measurements for each galaxy, e.g. the Petrosian and model system
\citep{stoughton2002sloan,Abazajian2009}. However, it is found
that the SDSS photometric measurements have not been optimized for
the galaxies with close neighbors \citep{Patton10}.

For J0848+1644, the galaxy A,B and C have the size $R_{90}$(the
circular aperture including 90 percent of the Petrosian flux)
8.87, 1.84 and 7.65 arcsec respectively(the dashed circles in Fig.
3). The underestimation of the size of galaxy B is because that
the SDSS algorithm make the photometric measurements for each
objects after masking the neighbors. So this small size further
leads to an underestimation of the relevant Petrosian magnitude of
galaxy B. To avoid such an over-subtraction, we reprocess the
photometry of J0848+1644 using the frame image from the SDSS data
archive. Rather than one by one, we make the photometry of their
members simultaneously. We take the Sersic profile for each member
galaxy and  make a combined fitting by using the 2D
\texttt{GALFIT} routine \citep{peng2002}.

The fitting results of the $r$-band image, including the total
magnitude ($m_r$), the S$\mathrm{\acute e}$rsic index ($n_s$), the
effective radius of the major axis ($R_e$), the ellipticity
($b/a$) and the position angle of the major axis along the triplet
alignment direction ($\theta$) are listed in Table \ref{tabel1}.
The residual image of the model fitting is shown in the bottom
right panel of Fig. \ref{Fig:plot1}. For comparison, we plot the
ellipse with $R_{a,90}$ that enclose 90\% model flux with solid
lines in the bottom left panel. It is clear that our fitting
results do not suffer from the over-masking problem, so they
should be more reliable to describe the photometric features of
these member galaxies.

Similarly, we also fit the SDSS $g$-band image and further derive
the $g-r$ color of each galaxy. The values listed in Table
\ref{tabel1} are already corrected for the Galactic extinction.

\section{Physical Properties} \label{prop}

In this Section, we measure and discuss the global properties of
J0848+1644 using the  photometric and spectroscopic data in the
above section, and summarize the results in Table \ref{tabel2}.

\subsection{Compactness and Configuration}

The compactness of a galaxy triplet is defined as a measurement of the
percentage of the system total area that is filled by the light of
member galaxies (\citealt{Duplancic2013}),

\begin{equation}\label{}
    S = \frac{\sum _{i=1}^{3} R_{90}^{2}}{R_m^2}
\end{equation}

\noindent where $R_{90}$ is the circle radius enclosing 90\% model
flux of the galaxy, $R_m$ is the radius of the minimum enclosing
circle that contains the geometric centers of all member galaxies
in the triplet. The $R_m$ of J0848+1644 is 12.2 kpc. Taking the
$R_{90}$ of galaxies in the $r-$band from our \texttt{GALFIT}
fitting ( $R_{c,90}$ in Table \ref{tabel1}), the compactness is $S =
2.48$, which is far larger than those of the O-sample, whose
median value is $\sim 0.05$.

It is worthy to mention that, for triplets, the apparent high compactness
may be contaminated by the projection effect. Considering the projection,  the apparent compactness
actually represents its upper limit. Anyway, according to the $S$ value, the J0848+1644
has a very high probability to be a highly compact system.

\cite{Agekyan1968} suggested an elegant method (AA-map) to
analyze the geometric configurations of the triplet systems.
They defined four types of configuration based on the
shape of the triangle formed by their members.

Obviously, J0848+1644 has a chain-like (alignment) configuration
(A-type) in the projected 2D AA-map. According to the
simulation result of \cite{Duplancic2015}, which counts the number
of mock triplets in each area of the 3D and 2D AA-map, J0848+1644
has 75\% probability to have a real 3D alignment configuration and
only 5\% probability to locate in the hierarchical region( H-type
).

\subsection{Environment}

In environmental studies, a galaxy is typically  defined as
isolated  if there is no neighbor galaxy with recessional
velocity difference $\Delta v \le$ 500 km s$^{-1}$  in a
particular projected distance, such as 1Mpc. For, J0848+1644, it
is certainly an isolated galaxy system, since its nearest
neighbor($M_r <$ -19.5) is 3.12 Mpc away.

To further quantify its isolation degree, following
\citet{Argudo2014}, we also calculate other two parameters.  One
is the number density of neighbor galaxies $\eta_{k,LSS}$, defined
as follows:
\begin{equation}\label{eq:eq1}
  \eta_{k,LSS} = \log\left( \frac {k-1}{V(r_k)}\right)
\end{equation}
where $V(r_k)=\frac 43 \pi r_k^3$ and $r_k$ is the projected
distance to the $k$th nearest neighbor of the large scale
structure (LSS), with $k$ equals to 5 or lower if there are not
enough neighbors in the field. According to the data of SDSS,
there are only four neighbor galaxies($M_r <$ -19.5) around with
radial velocity $\Delta v \le$ 500 km s$^{-1}$ in the volume of 5
Mpc projected distance. For J0848+1644, the large scale neighbor
galaxies number density $\eta_{4,LSS}$  is -2.10, which is much
less than the median value of the A-sample $\sim -1.4$. This
result even indicates that J0848+1644 locates in a lower density
environment than other isolated triplets.



Another parameter is about the tidal strength $Q$ on the primary
galaxy (here we use the central galaxy B instead) created by its
neighbors $i$ in the field
\begin{equation}\label{eq:eq2}
  Q = \log \left(\sum_i \frac {M_i}{M_P}\left(\frac {D_P}{d_{i}}\right)^3 \right)
\end{equation}
where $M_i$ is the stellar mass and $d_i$ is the projected
distance of the $i$th neighbor to the primary galaxy. $D_P =
2\alpha R_{90}$ is the diameter of the primary galaxy. It is
scaled by a factor $\alpha = 1.43$ to recover $D_{25}$
\citep{Argudo2014}. Obviously, the larger value of $Q$, the less
isolation the primary galaxy is.

For the triplet system, there are two measurements of the $Q$
parameter. $Q_{LSS}$ is generated by the neighbors of the triplet
up to 5 Mpc, but excludes its own companion galaxies. This is an
assessment of the isolation degree of the primary member galaxy in
the LSS environment. On the other hand, $Q_{Local}$ considers only
other two members of the triplet (galaxies A and C for
J0848+1644), which could represent the internal links within the
system itself.

Stellar masses of galaxies are estimated by the relationship
$log_{10}(M/L_r) = -0.306 + 1.097 (g-r)$ from \citet{bell2013}.
Thus, the $Q$ parameters of J0848+1644 are $Q_{LSS}= -5.70 $ and
$Q_{Local}= 1.29 $ respectively. It is worthy to be mentioned that
the the $Q_{triplet}$ may be overestimated because of the
projection effect.


\begin{figure}
   \centering
    \includegraphics[width=120mm]{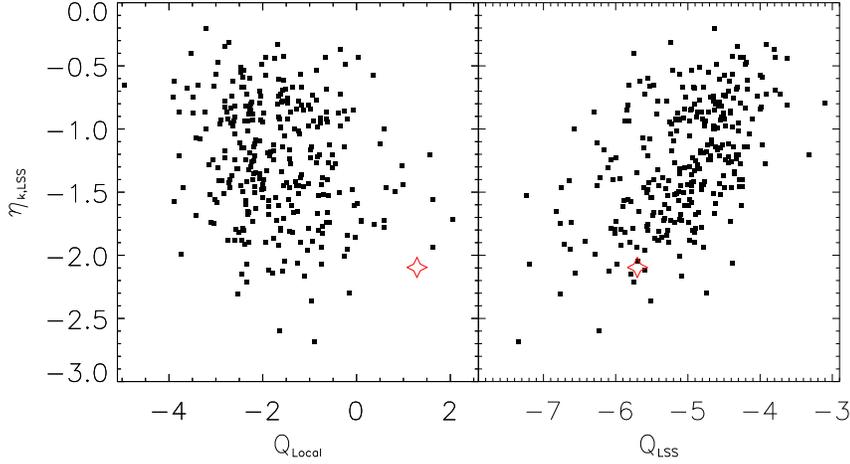}
   \caption{The environment parameters of triplet J0848+1644 (red stars) comparing
   with other isolated triplets in the A-sample (black dots). }
   \label{Fig:plot4}
\end{figure}

Comparing with galaxy triplets in A-sample as showing in Fig.
\ref{Fig:plot4}, $Q_{LSS}$ ( or $Q_{Local}$) of J0848+1644 is much
lower (or higher) than their median value -5.0 (or -2.0).
Additionally, the uniformed ratio of
$Q_{Local}/Q_{total}=Q_{Local}/(Q_{LSS}+Q_{Local})\simeq 1.0$. All
these values clearly identify that J0848+1644 is an extremely
compact and isolated system.

\subsection{Dynamics}

To characterize the dynamics of J0848+1644, we calculate  its
projected harmonic radius $R_H$, radial velocity dispersion
$\sigma_{v_r}$, dimensionless crossing time $H_0 t_c$ and virial
mass $M_{vir}$ respectively:
\begin{eqnarray}
  R_H &=& ( \frac {1}{N}\sum R_{ij}^{-1} )^{-1}, \\
  \sigma_{v_r}^2 &=& \frac {1}{N-1}\sum ( v_r - \langle v_r \rangle)^2, \\
  H_0 t_c &=& H_0 \pi R_H/(\sqrt{3}\sigma_{v_r}), \\
  M_{vir} &=& 3\pi N R_H\sigma_{v_r}^2 /(N-1)G.
\end{eqnarray}
%
where $R_{ij}$ to be the projections of galaxy-galaxy separation
and $N=3$ for the triplet.

Based on the $v_r$ results from the BFOSC spectra (see Table
\ref{tabel1} for details), we have $R_H=14.5 kpc$,
$\sigma_{v_r}=58.1 kms^{-1}$. Then, we can derive other two
parameters, $H_0 t_c =0.032$ and $\log(M_{vir}/M_{\bigodot})=
11.2$.

It is very interesting that the $M_{vir}$ of J0484+1644 is so
small, and even less than the total stellar mass of its member
galaxies. We are not surprised of this result, because for a
triplet system, the virial mass estimated by the radial velocity
dispersion might be strongly related to the viewing angle. For
example, if our line-of-sight happens to perpendicular to the
plane of three vectors of peculiar velocity of the triplet
members, the $\sigma_{v_r}$ could be deceased to zero. So, this
visual aspect reminds us that the $M_{vir}$ estimation of triplet
system is only a combination of the velocity dispersion
$\sigma_{v_r}$ and scale length $R_H$, which has the dimension of
mass \citep{Karachentseva2000}. It can not be considered
as the true mass of individual triplets, but should be worthwhile
in statistical analysis for a pretty large sample of triplets.

\subsection{Star formation rate}

It is believed that the galaxy interaction triggers star
formation. The galaxies in compact groups are shown to have
stronger star formation in comparison with those in fields
\citep{gomez2003}.  For galaxy B or C of J0848+1644, we see
significant H$\alpha$ emission line in their respective SDSS or
LAMOST spectrum, which indicates their current star forming
activities. Unfortunately, the LAMOST spectrum of galaxy C may be
seriously affected by unreliable flux calibration and causes an
unexpected stronger H$\alpha$ emission which may lead to the wrong
value of star formation rate (SFR).

Alternatively, we use the OII flux measured from the BFOSC
spectra of all three galaxies to calibrate their SFR, following
the relation of \citet{Kennicutt1998},
\begin{equation}
SFR(\text{M}_{\odot}\text{yr}^{-1}) \sim 1.4 \times 10^{-41} L
(\text{OII})(\text{ergs} \text{s}^{-1}),
\end{equation}
The SFR results, together with the $\log(SFR/M_*)$, are calculated
and listed in Table \ref{tabel1}. For galaxy B, these two values
are all consistent with those derived from the H$\alpha$ flux.
Then, we can sum up the SFR parameters for the whole triplet
system and list them in Table \ref{tabel2}.

From the result, galaxy A and B have lower star formation rate than C,
which causes OIII too weak, and H$\beta$ even too weak to
override stellar absorption (see Fig. \ref{Fig:plot2}).\citep{kennicutt1992}.




\begin{table}
\begin{center}
  \caption[]{Global physical parameters of J0848+1644}\label{tabel2}
  \smallskip
  \setlength{\tabcolsep}{14pt}
  \begin{threeparttable}
  \newcommand{\tabincell}[2]{\begin{tabular}{@{}#1@{}}#2\end{tabular}}
  \begin{tabular}{llrl}
  \hline\noalign{\smallskip}
  Parameter & Unit   &    J0848+1644    & NOTES \\
  \hline\noalign{\smallskip}
  $S$    &  &    2.48    &  Compactness parameter.\\
  $\eta_{4,LSS}$ &   &   -2.10   &  Projected number density.\\
  $Q_{LSS}$      &   &   -5.70   &  Tidal force effect of LSS.\\
  $Q_{Local}$    & &    1.29   &  Tidal force effect in the triplet.\\
  $R_H$ & kpc    &    14.5   & Projected Harmonic radii.\\
  $\sigma_{v_r}$ & km s$^{-1}$ &  58.1  & Radial velocity dispersion.\\
  $H_0 t_c$       &  &   0.032   &  Dimensionless crossing time.\\
  $\log(M_{vir}/M_{\bigodot})$  &  & 11.2 &  Virial mass. It may have strong projection effect. \\
  $SFR$  &  M$_{\odot}\cdot$ yr$^{-1}$  &  3.88      & Star formation rate      \\
  $\log(SFR/M_{\ast})$    &       & -10.51      &  sSFR       \\
\noalign{\smallskip}\hline
  \end{tabular}
  \end{threeparttable}
\end{center}
\end{table}

\section{DISCUSSION} \label{dis}

\subsection{Characteristics of J0848+1644}

The most distinguished feature of J0848+1644 is its compactness. The
parameter $S=2.48$ is located in the most compact end for almost
all current triplet samples. The lack of such compact triplets is
partly because of the sampling bias of redshift sky surveys (see
next subsection for details), and the investigation of J0848+1644
provides a valuable example to reveal the nature of such compact
few body system. Combining with other properties, the apparent
compactness seems to be real. All evidences point out that it
comes from a sufficient dynamical evolution, while the interaction
among member galaxies is still active.

From the projected 2D AA-map, J0848+1644 has a very high
probability to have an alignment configuration rather than a
hierarchical structure. It is believed  that different
configurations of the triplets may be reflecting different
dynamical stages of the system. The A-type configuration therefore
implies that it is a well formed system and its member galaxies
may be co-evolved for a long time. Then, the isolation
diagnosis brings out more information about the relations between
the J0848+1644 and its environment. Generally speaking, the
evolution of a galaxy may be affected by external influence when
the corresponding tidal force amounts to 1\% of the internal
binding force \citep{Athanassoula1984,Byrd1992}, that
corresponds to a tidal strength of $Q =-2$. Obviously, J0848+1644
has $Q_{LSS}$ (or $Q_{Local}$) more than three orders of magnitude
lower ( or higher ) than this level. That means this triplet has
extremely weak external influence from the LSS environment, and on
the contrast,  has very strong internal connection among its
members.

Additionally, the dimensionless crossing time $H_0t_c$ is the
ratio of the crossing time to the age of the universe and is a
convenient measurement of the system dynamical stage. For our
triplet, $H_0t_c=0.035$ is similar to the median value of the
O-sample (0.031) and not much longer than that of the K-sample
(0.019). Obviously it is much less than 1, which indicates the
sufficiency of dynamical evolution.  The direct consequence of
dynamical evolution is sinking the member galaxies into the deep
gravitational well of the central region of the dark matter halo,
and leading a small velocity dispersion since it is corresponding
to a smaller amount of mass of the innermost part of the halo. The
triplet J0848+1644 has a very small value of the radial velocity
dispersion. Although we have stated that it is probably due to the
projection, it is not conflict to the scenario of a well evolution
of this triplet.

The dynamical properties discussed above  also reflect  the
physical features of three member galaxies. Overall, members of
J0848+1644 are similar. They all have armless spherical shapes
with similar luminosity and color. This similarity is regarded as
another evidence of the long time co-evolution. All three
members show significant but not strong activities of current star
formation, that means the interaction may continue to play the
role of star forming trigger. It is worth to mention that the
colors and SFRs of these members are all near the boundary between
two classifications of galaxies. They are a little bit bluer, and
a little bit more active than non-star-forming galaxies. We are
not sure if this phenomenon is common for those very compact
systems at their final stage, or it is just an individual case for
the J0848+1644. So it is essential to increase the sample of this
kind of extreme compact triplet for the future researches.

\subsection{Importance of Spectroscopic Redshifts}

According to the spectroscopic observation of the BFOSC on the 2.16m
telescope, J0848+1644 is undoubtedly confirmed as a bounded
system, since the differences of $z_{spec}$ of its members are
much less than the typical radial velocity dispersions of triplet
systems. The reasons of why it was missed in previous samples, e.g.
the O-sample and the A-sample that all based on the SDSS, are the
fiber confliction effect and the relatively large uncertainties of
the $z_{phot}$ measurements. These two factors are all related to
the compactness of this triplet. Thus, there is a consequential
question that if these factors will seriously influence the study
of few body systems.

The fiber confliction effects are common in any redshift survey
using fiber spectroscopy. Taking the minimum fiber separation of
SDSS (55") as an typical angular distance, even for the low
redshift galaxies ($z\sim 0.1$), the closet pairs of galaxies that
can be observed in the same survey plate should have a distance
larger than 100 kpc. So usually we can not find and identify the
compact few body system by only using one scan of a fiber
spectroscopic survey data \cite[see][]{Shen2015}  for more
discussions. This is why \cite{omill2012} takes  $z_{phot}$ in
their searching of triplets. However, even taking the most
accurate measurements of $z_{phot}$, it has at least 5\%
uncertainties, which leads a $~1500kms^{-1}$ difference in radial
velocities at $z\sim 0.1$ and therefore is much larger than the
typical velocity dispersion of poor galaxy groups. Furthermore,
considering the neighboring effects in  the photometric
measurements(Section 2.2), just as the J0848+1644, the
uncertainties of radial velocity from $z_{phot}$ are even
enlarged.

For all of the above reasons, we believe that there is a
significant fraction of compact triplets of galaxies have not been
collected in current samples yet. The complement of such compact
systems is of great importance in understanding the final
evolution process of triplets, and also reveals some clues of the
connection between galaxy pairs and groups.

The discovery and identification of the triplet J0848+1644 also
declare that it is really important to carry out the supplemental
survey for current major galaxy redshift surveys , such as the
LAMOST, at least in the research field of compact few body systems
\citep{Shen2015}. Fortunately LAMOST has provided a good
opportunity to do such project. Although the LAMOST spectrum of galaxy C we used
here is occasionally noisier than most of ($\sim$92\%) DR1 spectra, it still provided
some vital information for our work, especially the obvious H$\alpha$
emission line which provides the reliable redshift rather than $z_{phot}$.

It is not easy to estimate how many compact
triplet systems could be found, before we have a complete
$z_{spec}$ sample. However, as a comparison, we used only
$z_{phot}$ to search the triplets in the A-sample though they all
have $z_{spec}$ measurements, and found more than $60\%$ real
triplets will be missed. This result strongly implies that
considerable amount of compact triplets will be found after some
effective supplemental surveys.

\section{SUMMARY} \label{sum}

We summarize the main results of this work as follows.

(1) The LAMOST spectra survey, which supplies new spectroscopic measurements to the main sample
galaxies without redshifts in SDSS due to the fiber collision, found the clue of a possible triplet
J0848+1644. It is further confirmed as a real triplet by the follow-up spectroscopic
observation of the BFOSC of the 2.16m telescope at Xinglong station
of NAOC.

(2) We find that the  J0848+1644
is an extremely compact isolated triplet, with alignment
configuration and very small radial velocity dispersion. The
member galaxies of this triplet have similar properties in their
shapes, colors and star formation rates. It gives an additional
example of the compact triplets, which are supposed to be co-evolved system rather
than the hierarchically forming structure.

(3) The compact systems like J0848+1644, e.g. the close pairs or poor groups,
are difficult to be gathered by the redshift survey
based on the fiber spectroscopy or using the current photometric redshift
technics.  The serendipitous discovery of J0848+1644  shows the importance of following projects
aiming at supplying the major redshift sky surveys, e.g. the LAMOST complementary galaxy survey \citet{Shen2015}.

\begin{acknowledgements}
We sincerely thank the anonymous referee whose suggestions
greatly helped us improve this paper. We are also grateful
to the kind staff at the Xinglong 2.16 m telescope for
their support during the observations.

This work is supported by the 973 Program" 2014 CB845705, Strategic
Priority Research Program The Emergence of Cosmological Structures" of the Chinese Academy
of Sciences (CAS; grant XDB09030200) and the National Natural Science Foundation of China
(NSFC) with the Project Number of 11390373, 11573050 and 11433003.

Funding for the creation and distribution of the SDSS archive has
been provided by the Alfred P. Sloan Foundation, the participating
institutions, NASA, the National Science Foundation, the US Department
of Energy, the Japanese Monbukagakusho and the Max
Plank Society. The SDSS website is http://www.sdss.org. SDSS is
managed by the Astrophysical Research Consortium for the participating
institutions.

The Large Sky Area Multi-Object Fiber Spectroscopic Telescope (LAMOST, now called the
Guoshoujing Telescope) is a National Major Scientific Project built by the Chinese Academy of
Sciences. Funding for the project has been provided by the National Development and Reform
Commission. LAMOST is operated and managed by the National Astronomical Observatories,
Chinese Academy of Sciences.

\end{acknowledgements}

\bibliographystyle{raa}
\bibliography{MS2519bibtex}

\begin{thebibliography}{30}
\providecommand{\natexlab}[1]{#1}
\providecommand{\selectlanguage}[1]{\relax}

\bibitem[{{Abazajian} et~al.(2009){Abazajian}, {Adelman-McCarthy},
  {Ag{\"u}eros} et~al.}]{Abazajian2009}
{Abazajian}, K.~N., {Adelman-McCarthy}, J.~K., {Ag{\"u}eros}, M.~A., et~al.
  2009, \apjs, 182, 543

\bibitem[{{Aceves}(2001)}]{Aceves2001}
{Aceves}, H. 2001, \mnras, 326, 1412

\bibitem[{{Agekyan} \& {Anosova}(1968)}]{Agekyan1968}
{Agekyan}, T.~A., \& {Anosova}, Z.~P. 1968, \sovast, 11, 1006

\bibitem[{{Anosova} et~al.(1992){Anosova}, {Kiseleva}, {Orlov}, \&
  {Chernin}}]{Anosova1992}
{Anosova}, Z.~P., {Kiseleva}, L.~G., {Orlov}, V.~V., \& {Chernin}, A.~D. 1992,
  \sovast, 36, 231

\bibitem[{{Argudo-Fern{\'a}ndez} et~al.(2014){Argudo-Fern{\'a}ndez}, {Verley},
  {Bergond} et~al.}]{Argudo2014}
{Argudo-Fern{\'a}ndez}, M., {Verley}, S., {Bergond}, G., et~al. 2014, \aap,
  564, A94

\bibitem[{{Argudo-Fern{\'a}ndez} et~al.(2015){Argudo-Fern{\'a}ndez}, {Verley},
  {Bergond} et~al.}]{Argudo2015}
{Argudo-Fern{\'a}ndez}, M., {Verley}, S., {Bergond}, G., et~al. 2015, ArXiv
  1504.00117

\bibitem[{{Athanassoula}(1984)}]{Athanassoula1984}
{Athanassoula}, E. 1984, \physrep, 114, 321

\bibitem[{{Bell} et~al.(2003){Bell}, {McIntosh}, {Katz}, \&
  {Weinberg}}]{bell2013}
{Bell}, E.~F., {McIntosh}, D.~H., {Katz}, N., \& {Weinberg}, M.~D. 2003, \apjs,
  149, 289

\bibitem[{{Byrd} \& {Howard}(1992)}]{Byrd1992}
{Byrd}, G.~G., \& {Howard}, S. 1992, \aj, 103, 1089

\bibitem[{{Chernin} \& {Mikkola}(1991)}]{Chernin1991}
{Chernin}, A.~D., \& {Mikkola}, S. 1991, \mnras, 253, 153

\bibitem[{Cui et~al.(2012)Cui, Zhao, Chu et~al.}]{Cui2012}
Cui, X.-Q., Zhao, Y.-H., Chu, Y.-Q., et~al. 2012, \raa, 12, 1197

\bibitem[{{Duplancic} et~al.(2015){Duplancic}, {Alonso}, {Lambas}, \&
  {O'Mill}}]{Duplancic2015}
{Duplancic}, F., {Alonso}, S., {Lambas}, D.~G., \& {O'Mill}, A.~L. 2015,
  \mnras, 447, 1399

\bibitem[{{Duplancic} et~al.(2013){Duplancic}, {O'Mill}, {Lambas}, {Sodr{\'e}},
  \& {Alonso}}]{Duplancic2013}
{Duplancic}, F., {O'Mill}, A.~L., {Lambas}, D.~G., {Sodr{\'e}}, L., \&
  {Alonso}, S. 2013, \mnras, 433, 3547

\bibitem[{{G{\'o}mez} et~al.(2003){G{\'o}mez}, {Nichol}, {Miller}
  et~al.}]{gomez2003}
{G{\'o}mez}, P.~L., {Nichol}, R.~C., {Miller}, C.~J., et~al. 2003, \apj, 584,
  210

\bibitem[{{Karachentsev}(1990)}]{karachentsev1990}
{Karachentsev}, I.~D. 1990, in NASA Conference Publication, \emph{NASA
  Conference Publication}, vol. 3098, edited by J.~W. {Sulentic}, W.~C. {Keel},
  \& C.~M. {Telesco}, 3--17

\bibitem[{{Karachentsev} \& {Karachentseva}(1981)}]{Karachentsev1981}
{Karachentsev}, I.~D., \& {Karachentseva}, V.~E. 1981, Astrofizika, 17, 5

\bibitem[{{Karachentsev} et~al.(1988){Karachentsev}, {Karachentsev}, \&
  {Lebedev}}]{Karachentsev1988}
{Karachentsev}, V.~E., {Karachentsev}, I.~D., \& {Lebedev}, V.~S. 1988,
  Astrofizicheskie Issledovaniia Izvestiya Spetsial'noj Astrofizicheskoj
  Observatorii, 26, 42

\bibitem[{{Karachentseva} \& {Karachentsev}(1982)}]{Karachentseva1982}
{Karachentseva}, V.~E., \& {Karachentsev}, I.~D. 1982, Astrofizika, 18, 5

\bibitem[{{Karachentseva} \& {Karachentsev}(1983)}]{Karachentseva1983}
{Karachentseva}, V.~E., \& {Karachentsev}, I.~D. 1983, Astrofizika, 19, 613

\bibitem[{{Karachentseva} \& {Karachentsev}(2000)}]{Karachentseva2000}
{Karachentseva}, V.~E., \& {Karachentsev}, I.~D. 2000, Astronomy Reports, 44,
  501

\bibitem[{{Karachentseva} et~al.(1979){Karachentseva}, {Karachentsev}, \&
  {Shcherbanovsky}}]{Karachentseva1979}
{Karachentseva}, V.~E., {Karachentsev}, I.~D., \& {Shcherbanovsky}, A.~L. 1979,
  Astrofizicheskie Issledovaniia Izvestiya Spetsial'noj Astrofizicheskoj
  Observatorii, 11, 3

\bibitem[{{Kennicutt}(1992)}]{kennicutt1992}
{Kennicutt}, R.~C., Jr. 1992, \apj, 388, 310

\bibitem[{{Kennicutt}(1998)}]{Kennicutt1998}
{Kennicutt}, R.~C., Jr. 1998, \araa, 36, 189

\bibitem[{{Luo} et~al.(2015){Luo}, {Zhao}, {Zhao} et~al.}]{Luo2015}
{Luo}, A.-L., {Zhao}, Y.-H., {Zhao}, G., et~al. 2015, ArXiv: 1505.01570

\bibitem[{{O'Mill} et~al.(2012){O'Mill}, {Duplancic}, {Garc{\'{\i}}a Lambas},
  {Valotto}, \& {Sodr{\'e}}}]{omill2012}
{O'Mill}, A.~L., {Duplancic}, F., {Garc{\'{\i}}a Lambas}, D., {Valotto}, C., \&
  {Sodr{\'e}}, L. 2012, \mnras, 421, 1897

\bibitem[{{Patton} et~al.(2011){Patton}, {Ellison}, {Simard}, {McConnachie}, \&
  {Mendel}}]{Patton10}
{Patton}, D.~R., {Ellison}, S.~L., {Simard}, L., {McConnachie}, A.~W., \&
  {Mendel}, J.~T. 2011, \mnras, 412, 591

\bibitem[{{Peng} et~al.(2002){Peng}, {Ho}, {Impey}, \& {Rix}}]{peng2002}
{Peng}, C.~Y., {Ho}, L.~C., {Impey}, C.~D., \& {Rix}, H.-W. 2002, \aj, 124, 266

\bibitem[{{Shen} et~al.(2015){Shen}, {Argudo-Fern$\mathrm{\acute a}$ndez},
  {Chen} et~al.}]{Shen2015}
{Shen}, S.-Y., {Argudo-Fern$\mathrm{\acute a}$ndez}, M., {Chen}, L., et~al.
  2015, submitted to \raa

\bibitem[{Stoughton et~al.(2002)Stoughton, Lupton, Bernardi
  et~al.}]{stoughton2002sloan}
Stoughton, C., Lupton, R.~H., Bernardi, M., et~al. 2002, The Astronomical
  Journal, 123, 485

\bibitem[{{Zheng} et~al.(1993){Zheng}, {Valtonen}, \& {Chernin}}]{Zheng1993}
{Zheng}, J.-Q., {Valtonen}, M.~J., \& {Chernin}, A.~D. 1993, \aj, 105, 2047

\end{thebibliography}

\end{document}